# Observation of Acoustic Valley Vortex States and Valley-Chirality Locked Beam Splitting

Liping Ye,[1] Chunyin Qiu,[1,*] Jiuyang Lu,[4] Xinhua Wen,[1] Yuanyuan Shen,[1] Manzhu Ke,[1] Fan Zhang,[3] and Zhengyou Liu[1,2]

[1] Key Laboratory of Artificial Micro- and Nano-structures of Ministry of Education and School of Physics and Technology, Wuhan University, Wuhan 430072, China
[2] Institute for Advanced Studies, Wuhan University, Wuhan 430072, China
[3] Department of Physics, University of Texas at Dallas, Richardson, Texas 75080, USA
[4] Department of Physics, South China University of Technology, Guangzhou 510641, China

**Abstract**: We report an experimental observation of the classical version of valley polarized states in a two-dimensional hexagonal sonic crystal, where the inversion-symmetry breaking of scatterers induces an omnidirectional frequency gap. The acoustic valley states, which carry specific linear momenta and orbital angular momenta, were selectively excited by external Gaussian beams and conveniently confirmed by the pressure distribution outside the crystal, according to the criterion of momentum conservation. The vortex nature of such intriguing crystal states was directly characterized by scanning the phase profile inside the crystal. In addition, we observed a peculiar beam splitting phenomenon, in which the separated beams are constructed by different valleys and locked to the opposite vortex chirality. The exceptional sound transport, encoded with valley-chirality locked information, may serve as the basis of designing conceptually novel acoustic devices with unconventional functions.

Sound is one of the most common waves in daily life. However, sound wave is not easy to control by external fields, e.g., electric and magnetic fields, since it lacks intrinsic degrees of freedom (DOFs) like the charge and spin in electrons. As a fundamental property of waves, sound can be scattered or diffracted when suffering inhomogeneity in the propagation path. This offers an efficient route of steering sound by artificial structures, e.g., sonic crystals (SCs) [1-4], acoustic metamaterials [5-8] and metasurfaces [9-12]. By utilizing these unnatural media, the propagation of sound can be tailored in unprecedented ways, such as negative refraction [1-3,13-15], super-resolution imaging [3,16,17], cloaking [18-20], and abnormal wavefront shaping [9-12]. Recently, the acoustic structures have also been demonstrated to be good platforms in exploring topological physics predicted originally in electronic systems [21-28].

In condensed matter systems, the property of valley electrons has sparked extensive interest in recent years [29-43]. The discrete valley index, labeling the degenerate energy extrema in momentum space, can be treated as a new quantum DOF other than charge and spin when the intervalley scattering is negligibly weak. Numerous fascinating phenomena associated with valley-contrasting properties have been studied, such as valley filters and valley Hall effects, which are paving the road for the applications of valleytronics (e.g., in quantum computing and information processing). Inspired by the concept of valley DOF in electronic states, recently, Lu *et al.* have theoretically studied the valley states of acoustic waves in two-dimensional (2D) SCs, and revealed that such states carry notable vortex signatures [44]. Particularly, the acoustic valley (AV) states locked to specific vortex chirality can be independently accessed by external sound according to their linear or angular momenta. Therefore, analog to the electronic case, the additional valley DOF in the artificially designed SC structure, associated with the valley-chirality locking property, could also construct a good carrier of information and thus enables a brand new manner to control sound.

In this Letter, we present the first experimental observation of the AV vortex states in a 2D hexagonal SC. Valley-selective excitation has been realized by an external Gaussian beam at a particular incident angle, and confirmed by the Fourier spectrum of the pressure field outside the SC sample, together with an elaborate characterization of the vortex chirality through detecting the phase profile around the vortex core. Furthermore, we demonstrate a unique beam splitting behavior, where two spatially separated beams are locked to different vortex chirality. The experimental data agree fairly well with the full-wave simulations performed by COMSOL Multiphysics. As such, by exciting the AV states we have experimentally demonstrated a novel manipulation on sound: making vortex arrays and controlling their chirality according to the moving directions. In addition, as a natural property of the chiral phased waves, the sound vortices carry orbital angular momentum (OAM) inherently [45], and thus enable many tantalizing applications, e.g., to generate mechanical torques on the trapped objects by delivering the OAM of sound to matter [46-52].

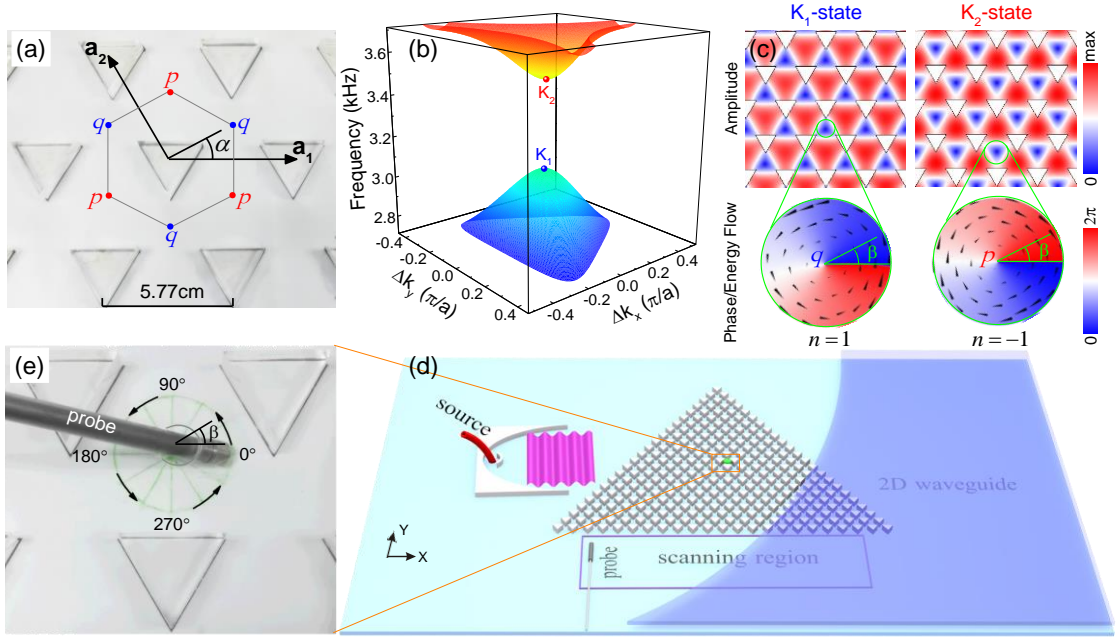

FIG. 1. (a) A local photograph of the SC sample made of a hexagonal array of regular triangle plexiglass rods, with $p$ and $q$ labeling two inequivalent lattice centers. (b) The lowest two bands plotted around the $K$ point. (c) Eigenfields for the valley states $K_1$ and $K_2$. The arrow indicates the orientation of vortex-like energy flow. (d) Schematic of the experimental setup. All measurements are performed inside a 2D air-filled waveguide formed by two parallel plexiglass plates. (e) Phase scanning

along a circular loop by counterclockwise rotating the probe, with $\beta$ labeling the azimuthal angle.

As shown in Fig. 1(a), the SC is built by a 2D hexagonal array of regular triangle plexiglass rods. The length of lattice vectors $a = |\mathbf{a}_1| = |\mathbf{a}_2| = 5.77$ cm and the side length of each rod $l = 3.0$ cm. These dimensions are carefully designed to allow a field scanning inside the unit cell, meanwhile considering a trade-off with the total sample size limited by our measuring platform. (Principally, a sample containing more unit cells is preferred to reducing the finite size effect.) The orientation DOF of the anisotropic rod, depicted by the rotation angle $\alpha$, offers a flexible control of the band gap [53]. Specifically, here $\alpha = 30^\circ$ is utilized to open a sizable omnidirectional band gap between the lowest two bands [Fig. 1(b)]. Below we focus on the AV states $K_1$ (3.06 kHz) and $K_2$ (3.49 kHz) that locate at the corner point $K$ of the first Brillouin zone. Similar to the orbital motion of valley electrons, each state is featured by a hexagonal array of sound vortices [Fig. 1(c)], centered at the equivalent lattice centers $q$ (or $p$) and rotated counterclockwise (or clockwise). Intriguingly, the AV vortex states carry quantized phase winding numbers ($n = \pm 1$), owing to the threefold rotation symmetry of the system [28,44]. The counterparts at the inequivalent valley $K'$, possess invariant vortex cores (with zero amplitudes) but opposite chirality, as required by time-reversal symmetry.

It has been pointed out that [44], different from the electronic case where the excitation of the valley polarized states resorts to additional fields [54-60], the AV states can be directly accessed by sound stimuli, either by Gaussian beams based on the principle of momentum conservation, or by point-like chiral sources according to the azimuthal selection rule. Here we employ the first approach since a Gaussian beam can be prepared easily in experiment. Figure 1(d) shows our experimental setup. The sample contains 276 rods in total and has a shape of regular triangle with side lengths 130 cm. Its surface normal is selected along the $\Gamma M$ direction to avoid intervalley scattering at the SC boundary [44]. Thanks to the macroscopic characteristic of the system, the triangular plexiglass rods, fabricated by laser-cutting, can be orientated and arranged into the hexagonal lattice precisely, which enables the valley transport nearly free of short range scattering inside the SC. To guarantee the experimental system of 2D nature, the whole structure is positioned in an air-filled planar sound waveguide (of height 1.2 cm) formed by two plexiglass plates, which allows only the propagation of the fundamental waveguide mode for the concerned wavelength (>9.3 cm). The point-like sound signal, launched from a narrow tube with inner diameter ~0.8 cm, is transferred into a Gaussian beam when it is reflected by a carefully designed parabolic concave mirror [61]. The width of the Gaussian beam is controlled by the geometry of the concave mirror. The pressure field inside or outside the SC sample can be scanned by a movable microphone of diameter ~0.7 cm (B&K Type 4187). Specifically, to confirm the vortex chirality of the AV states, a small hole (of diameter ~1.1 cm) is drilled through the top plate, and the probe microphone is obliquely inserted into the waveguide. As illustrated in Fig. 1(e), by rotating the probe (at an angular step of $30^\circ$) one can detect the pressure response along a circular contour centered at a given lattice center (e.g., $p$ point for the $K_2$ state here). Finally, the sound signal is analyzed by a multi-analyzer system (B&K Type 3560B), from which the phase and amplitude of the pressure field can be extracted simultaneously.

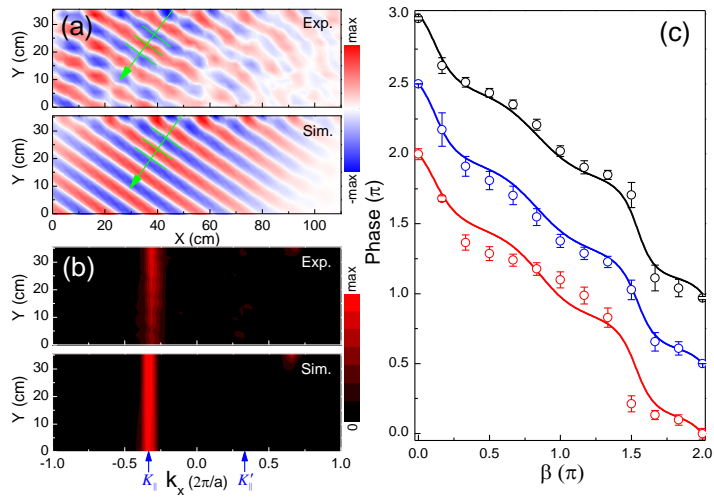

FIG. 2. (a) Pressure patterns measured and simulated outside the sample, stimulated by an obliquely incident Gaussian beam at the $K_2$ frequency. The green arrows

indicate the propagation of sound wavefronts. (b) The corresponding spatial Fourier spectra performed along the *x* direction. (c) Measured (circles) and simulated (lines) phase distributions for three different circular loops encircling the equivalent *p* points. The error bars represent the standard deviation of the measurement.

To experimentally stimulate the $K_2$ state, we launch a Gaussian beam of width ~23cm onto the SC sample. Specifically, the incident angle $\theta \approx 34°$ satisfies $k_0 \sin\theta = K_\parallel$, where $k_0$ is the wavenumber in air space and $K_\parallel = 2\pi/3a$ is a projection of the $K$ point to the sample boundary. By this incidence, the $K_2$ state is anticipated to be well excited owing to the momentum conservation parallel to the SC boundary. Instead of directly scanning the whole field inside the sample, an easier task is carried out to validate the valley selection, i.e., detecting the sound signal leaked out. For example, we consider the spatial region earmarked by the rectangle in Fig. 1(d). As displayed visually by Fig. 2(a), the experimentally measured wavefront pattern agrees well with the COMSOL simulation. A nearly pure excitation of the $K_2$ state can be further checked from the corresponding Fourier spectra performed parallel to the horizontal sample boundary [Fig. 2(b)]. As predicted by the theory, the experimental data shows a bright stripe at $k_x = K_\parallel$, where the momentum broadening stems mostly from the finite size effect. This is in striking contrast to the position $k_x = K'_\parallel$, a projection of the $K'$ point. Figure 2(c) exemplifies the phase profiles for three circular loops located in different unit cells. In all cases, the experimental measurements, in good agreement with the simulations, exhibit a phase reduction of $2\pi$ over the loop, another critical signature for the excitation of the $K_2$ state. Note that the quantized phase winding number ($n = -1$) is easy to capture experimentally, since the vortex singularity is generic and structurally stable in real space [45]. Such immunity to noise perturbation is appealing in practical applications. Similarly, the $K_1$ state (with $n = 1$) can also be excited and detected independently (see *Supplementary Information*).

The above study states that, by selectively exciting a single valley state, one can realize a hexagonal lattice of sound vortices. Their chirality can be controlled by the incident frequency and direction of the Gaussian beam. As such, the distinguishable valley signature, i.e., valley-chirality locking, enables a new DOF to manipulate sound. This is greatly meaningful for scalar acoustics that lacks intrinsic DOFs like charge and spin. In addition, as a fundamental morphology of sound profile, the vortex matter of sound (imprinted with nonzero OAM) is attracting fast growing attention due to its promising applications [46-52]. Usually, a sound vortex is realized by an array of active sound sources with elaborate phase lags [46-51,62-64]. Recently, artificial structures, e.g., spiral gratings [65-67] and metamaterials [68,69], have been proposed to make sound vortices. Strikingly different from these passive schemes, which produce only a single vortex that travels spirally along its axis, the current method generates a compact array of vortices simultaneously, and the energy transports in the 2D lattice plane.

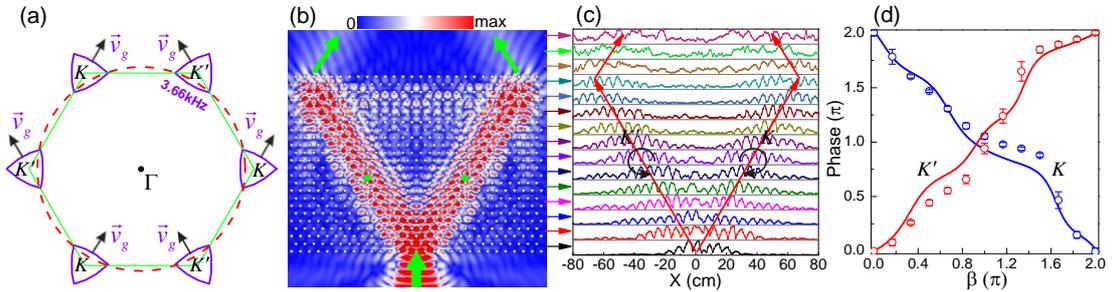

FIG. 3. (a) Trigonal warping effect of the EFC (purple solid lines), illustrated by a frequency slightly above $K_2$. The red dashed line indicates the circular EFC of air at the same frequency. (b) Amplitude distribution of the pressure field simulated for a narrow Gaussian beam incident normally from the bottom. (c) Experimentally measured pressure amplitudes along the labeled horizontal lines. The red arrows guide the propagation of the split sub-wave-packets. (d) Phase profile (circles) scanned along two circular loops located on the left- and right-moving beams [see green points in (b)], together with the numerical data (lines) for comparison.

Below we demonstrate another fascinating manipulation of the AV vortex states,

i.e., stimulating the valleys $K$ and $K'$ simultaneously and separating them in different spatial regions. This valley-dependent beam splitting stems essentially from a trigonal warping effect of the band structure [44,70]. As exemplified by the EFCs of 3.66 kHz [Fig. 3(a)], if an incident beam covers a broad range of momentum distribution, the forward-moving states around the $K$ and $K'$ points will be excited at the same time. Considering the finite lengths of the trigonal EFCs in momentum space, the beams constructed by these states must be finite widths, and separate as they propagate because of the different group velocities [$\vec{v}_g$ depicted by the arrows in Fig. 3(a)]. This is displayed clearly in Fig. 3(b). Interestingly, each beam carries a chiral feature of the corresponding valley, since the vortex-like field profile remains even for the state off the band edge. To experimentally confirm the valley-chirality locked beam splitting effect, we have fabricated a rectangular SC sample of size 164x125 cm$^2$ (made of 684 rods, as used in the simulation). The size of the reflective concave mirror is reduced to generate a narrower Gaussian beam (of width ~10.0 cm). By inserting the detecting probe inside the SC, we have experimentally measured the pressure distributions (at 3.66 kHz) along several equidistant horizontal lines. As observed in Fig. 3(c), the amplitude distribution shows a clear separation of the wave-packet as the sound travels forward, in which the traces of the sub-wave-packets agree well with the theoretical prediction. The main contribution of Bloch states for each beam has been checked further by performing Fourier transform (along the *x* direction) for the corresponding sub-wave-packet: as specified in Fig. 3(c), the $K$ and $K'$ valleys are responsible for the right- and left-moving beams, respectively. To identify the vortex chirality carried by these two beams, we have scanned the phase profiles for two circular loops that locate separately on the two beams. The experimental data in Fig. 3(d), consistent with the COMSOL simulations, display obvious anticlockwise ($n=1$) and clockwise ($n=-1$) phase growths for the left- and right-moving beams, respectively. In some sense, this peculiar beam splitting behavior resembles the valley Hall effect of electronic systems: the deflection of each sound beam is linked with particular vortex chirality. Interestingly, both split beams travel in a 'wrong' way when they are refracted out from the sample: the left-moving beam deflects rightward and the right-moving beam deflects leftward. This anomalous propagation of sound, so-called negative refraction, has sparked intense interest in past decades [1-3,13-15], which can be explained straightforwardly from the shape of EFC.

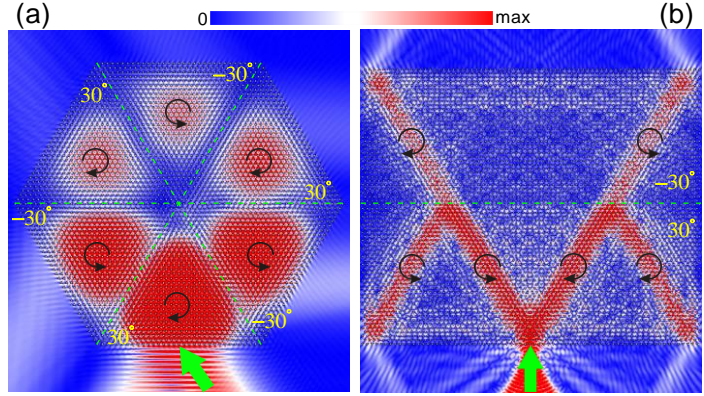

FIG. 4. OAM-reversed valley transports in the samples constructed by $\alpha$-inverted SCs (here $\alpha = \pm 30^\circ$), simulated for (a) the selective excitation of the $K_2$ state and (b) the beam splitting behavior.

It is worth mentioning that, the operating frequency can be continuously tailored by mechanically rotating the triangular scatterers. Particularly, if the rotation angle $\alpha$ is inverted, the vortex chirality can even be switched without any change of the incident wave, considering a mirror operation of the system. Several fancy sound transports can be imagined further if such $\alpha$-inverted SCs are combined together. For example, Fig. 4(a) shows an alternate excitation of the valley vortices with the opposite OAMs in the neighboring SC blocks with $\alpha = \pm 30^\circ$. The whole sample is well ignited because the momentum conservation is fulfilled perfectly at each SC interface. Similar idea can be extended to the valley-chirality locked beam splitting. As demonstrated in Fig. 4(b), the OAMs of the sound vortices are reversed separately once the beams traverse the SC interface (where the reflection can be remarkably reduced by adding a few SC layers with gradient $\alpha$ distribution). Note that so far we have been focusing on the AV states of the lowest two bands. The AV states located on the higher bands, which accommodate more fine features, also deserve special attention. For example, the valley state on the third band is characterized by a pair of oppositely-rotated vortices in a single unit cell, centered at the *p* and *q* points respectively. Numerical and experimental data are provided in the *Supplemental Information*.

    In conclusion, the universal valley physics has been successfully validated in the acoustic system (associated with reconfigurable operation frequencies and vortex chirality by mechanically rotating the triangular rods). This may stimulate extensive interest in the exploration of the valley-dependent phenomena in various artificial crystals and lead to novel manipulations on the corresponding classical waves. In addition to the implications in fundamental researches, such compact sound vortex array with controllable chirality, which is unattainable through the currently existing approaches [62-69], could be potentially useful for patterning and rotating objects without contact, given the interaction of sound with matter [44,46-52,64,67,69].

**Acknowledgements**

The authors thank C. T. Chan for fruitful discussions. This work is supported by the National Basic Research Program of China (Grant No. 2015CB755500); National Natural Science Foundation of China (Grant Nos. 11674250, 11374233, 11534013, and 11547310); Postdoctoral innovation talent support program (BX201600054). F. Zhang was supported by the UT-Dallas research enhancement funds.

**References:**

[1] S. Yang, J. H. Page, Z. Liu, M. L. Cowan, C. T. Chan, and P. Sheng, Phys. Rev. Lett. **93**, 024301 (2004).

[2] M. Lu, C. Zhang, L. Feng, J. Zhao, Y. Chen, Y. Mao, J. Zi, Y. Zhu, S. Zhu, and N. Ming, Nat. Mater. **6**, 744–748 (2007).

[3] A. Sukhovich, B. Merheb, K. Muralidharan, J. O. Vasseur, Y. Pennec, P. A. Deymier, and J. H. Page, Phys Rev Lett. **102**, 154301 (2009).

[4] X. Zhang and Z. Liu, Phys. Rev. Lett. 101, 264303 (2008).

[5] Z. Liu, X. Zhang, Y. Mao, Y. Y. Zhu, Z. Yang, C. T. Chan, and P. Sheng, Science **289**, 1734 (2000).

[6] N. Fang, D. Xi, J. Xu, M. Ambati, W. Srituravanich, C. Sun, and X. Zhang, Nat. Mater. **5**, 452 (2006).

[7] M. Yang, G. Ma, Z. Yang, and P. Sheng, Phys. Rev. Lett. **110**, 134301 (2013).

[8] S. H. Lee, C. M. Park, Y. M. Seo, Z. G. Wang, and C. K. Kim, Phys. Rev. Lett. **104**, 054301 (2010).

[9] Y. Li, B. Liang, Z. Gu, X. Zou, and J. Cheng, Sci. Rep. **3**, 2546 (2013).

[10] J. Zhao, B. Li, Z. Chen, and C. W. Qiu, Sci. Rep. **3**, 2537 (2013).

[11] K. Tang, C. Qiu, M. Ke, J. Lu, Y. Ye, and Z. Liu, Sci. Rep. **4**, 6517 (2014).

[12] Y. Xie, W. Wang, H. Chen, A. Konneker, B.-I. Popa, and S. A. Cummer, Nat. Commun. **5**, 5553 (2014).

[13] Z. Liang, and J. Li, Phys. Rev. Lett. **108**, 114301 (2012).

[14] J. Christensen, and F. J. Garcia de Abajo, Phys. Rev. Lett. **108**, 124301 (2012).

[15] V. M. Garcia-Chocano, J. Christensen, and J. Sanchez-Dehesa, Phys. Rev. Lett. **112**, 144301 (2014).

[16] J. Li, L. Fok, X. Yin, G. Bartal, and X. Zhang, Nat. Mater. **8**, 931–934 (2009).

[17] J. Zhu, J. Christensen, J. Jung, L. Martin-Moreno, X. Yin, L. Fok, X. Zhang, and F. J. Garcia-Vidal, Nat. Phys. **7**, 52–55 (2011).

[18] S. Zhang, C. Xia, and N. Fang, Phys. Rev. Lett. **106**, 024301 (2011).

[19] B.-I. Popa, L. Zigoneanu, and S. A. Cummer, Phys. Rev. Lett. **106**, 253901 (2011).

[20] L. Zigoneanu, B.-I. Popa, and S. A. Cummer, Nat. Mater. **13**, 352 (2014).

[21] Z. Yang, F. Gao, X. Shi, X. Lin, Z. Gao, Y. Chong, and B. Zhang, Phys. Rev. Lett. **114**, 114301 (2015).

[22] P. Wang, L. Lu, and K. Bertoldi, Phys. Rev. Lett. **115**, 104302 (2015).

[23] M. Xiao, G. Ma, Z. Yang, Z. Q. Zhang, and C. T. Chan, Nat. Phys. **11**, 240 (2015).

[24] V. Peano, C. Brendel, M. Schmidt, and F. Marquardt, Phys. Rev. X 5, 031011 (2015).

[25] C. He, X. Ni, H. Ge, X. Sun, Y. Chen, M. Lu, X. Liu, and Y. Chen, Nat. Phys. **12**, 1124 (2016).

[26] Y. Peng, C. Qin, D. Zhao, Y. Shen, X. Xu, M. Bao, H. Jia, and X. Zhu, Nat. Commun. **7**, 13368 (2016).

[27] J. Lu, C. Qiu, L. Ye, X. Fan, M. Ke, F. Zhang, and Z. Liu, Nat. Phys. 2016, DOI: 10.1038/NPHYS3999.

[28] C. Brendel, V. Peano, O. Painter, and F. Marquardt, arXiv:1607.04321 (2016).

[29] A. Rycerz, J. Tworzydlo, and C. W. J. Beenakker, Nat. Phys. **3**, 172 (2007).

[30] D. Xiao, W. Yao, and Q. Niu, Phys. Rev. Lett. **99**, 236809, (2007).

[31] R. V. Gorbachev, J. C. W. Song, G. L. Yu, A. V. Kretinin, F. Withers, Y. Cao, A. Mishchenko, I. V. Grigorieva, K. S. Novoselov, L. S. Levitov, and A. K. Geim, Science **346**, 448 (2014).

[32] D. Xiao, G. Liu, W. Feng, X. Xu, and W. Yao, Phys. Rev. Lett. **108**, 196802 (2012).

[33] K. F. Mak, K. L. McGill, J. Park, and P. L. McEuen, Science **344**, 1489 (2014).

[34] X. Xu, W. Yao, D. Xiao, and T. F. Heinz, Nat. Phys. **10**, 343-350 (2014).

[35] H. Pan, X. Li, F. Zhang, and S. A. Yang, Phys. Rev. B **92**, 041404(R) (2015).

[36] G. W. Semenoff, V. Semenoff, and F. Zhou, Phys. Rev. Lett. **101**, 087204 (2008).

[37] I. Martin, Y. M. Blanter, and A. F. Morpurgo, Phys. Rev. Lett. **100**, 036804 (2008).

[38] F. Zhang, J. Jung, G. A. Fiete, Q. Niu, and A. H. MacDonald, Phys. Rev. Lett. **106**, 156801 (2011).

[39] Z. Qiao, J. Jung, Q. Niu, and A. H. Macdonald, Nano Lett. **11**, 3453-3459 (2011).

[40] F. Zhang, A. H. MacDonald, and E. J. Mele, Proc. Natl Acad. Sci. USA **110**, 10546-10551 (2013).

[41] A. Vaezi, Y. Liang, D. H. Ngai, L. Yang, and E.-A. Kim, Phys. Rev. X **3**, 021018 (2013).

[42] L. Ju, Z. Shi, N. Nair, Y. Lv, C. Jin, J. V. Jr, C. Ojeda-Aristizabal, H. A. Bechtel, M. C. Martin, A. Zettl, J. Analytis, and F. Wang, Nature **520**, 650-655 (2015).

[43] J. Li, K. Wang, K. McFaul, Z. Zern, Y. Ren, K. Watanabe, T. Taniguchi, Z. Qiao, and J. Zhu, Nat. Nanotech. **11**, 1060 (2016).

[44] J. Lu, C. Qiu, M. Ke, and Z. Liu, Phys. Rev. Lett. **116**, 093901 (2016).

[45] J. F. Nye and M. V. Berry, Proc. R. Soc. A **336**, 165 (1974).

[46] K. Volke-Sepulveda, A. O. Santillan, and R. R. Boullosa, Phys. Rev. Lett. **100**, 024302 (2008).

[47] K. D. Skeldon, C. Wilson, M. Edgar, and M. J. Padgett, New J. Phys. **10**, 013018 (2008).

[48] A. O. Santillán, and K. A. Volke-Sepúlveda, Am. J. Phys. **77**, 209 (2009).

[49] C. E. M. Demore, Z. Yang, A. Volovick, S. Cochran, M. P. MacDonald, and G. C. Spalding, Phys. Rev. Lett. **108**, 194301 (2012).

[50] A. Anhauser, R. Wunenburger, and E. Brasselet, Phys. Rev. Lett. **109**, 034301 (2012).

[51] Z. Y. Hong, J. Zhang, and B. W. Drinkwater, Phys. Rev. Lett. **114**, 214301 (2015).

[52] L. K. Zhang and P. L. Marston, Phys. Rev. E **84**, 065601 (2011).

[53] J. Lu, C. Qiu, S. Xu, Y. Ye, M. Ke, and Z. Liu, Phys. Rev. B **89**, 134302 (2014).

[54] K. Takashina, Y. Ono, A. Fujiwara, Y. Takahashi, and Y. Hirayama, Phys. Rev. Lett. **96**, 236801 (2006).

[55] O. Gunawan, Y. P. Shkolnikov, K. Vakili, T. Gokmen, E. P. De Poortere, and M. Shayegan, Phys. Rev. Lett. **97**, 186404 (2006).

[56] Z. Zhu, A. Collaudin, B. Fauque, W. Kang, and K. Behnia, Nat. Phys. **8**, 89-94 (2011).

[57] D. MacNeill, C. Heikes, K. F. Mak, Z. Anderson, A. Kormányos, V. Zólyomi, J. Park, and D. C. Ralph, Phys. Rev. Lett. **114**, 037401 (2015).

[58] K. F. Mak, K. He, J. Shan, and T. F. Heinz, Nat. Nanotechnol. **7**, 494 (2012).

[59] H. Zeng, J. Dai, W. Yao, D. Xiao, and X. Cui, Nat. Nanotechnol. 7, 490 (2012).

[60] T. Cao, G. Wang, W. Han, H. Ye, C. Zhu, J. Shi, Q. Niu, P. Tan, E. Wang, B. Liu, and J. Feng, Nat. Commun. **3**, 887 (2012).

[61] J. Lu, C. Qiu, S. Xu, Y. Ye, M. Ke, and Z. Liu, Phys. Rev. B **89**, 134302 (2014).

[62] B. T. Hefner and P. L. Marston, J. Acoust. Soc. Am. **106**, 3313 (1999).

[63] R. Marchiano, and J-L. Thomas, Phys. Rev. E **71**, 066616 (2005).

[64] D. Baresch, J.-L. Thomas, and R. Marchiano, Phys. Rev. Lett. **116**, 024301 (2016).

[65] N. Jimenez, V. J. Sanchez-Morcillo, R. Pico, L. M. Garcia-Raffi, V. Romero-Garcia, K. Staliunas, Physics Procedia. **70**, 245-248 (2015).

[66] X. Jiang, J. J. Zhao, S. L. Liu, B. Liang, X. Zou, J. Yang, C. Qiu, and J. C. Cheng, Appl. Phys. Lett. **108**, 203501 (2016).

[67] T. Wang, M. Ke, W. Li, Q. Yang, C. Qiu, and Z. Liu, Appl. Phys. Lett. **109**, 123506 (2016).

[68] X. Jiang, Y. Li, B. Liang, J. Cheng, and L. Zhang, Phys. Rev. Lett. **117**, 034301 (2016).

[69] L. Ye, C. Qiu, J. Lu, K. Tang, H. Jia, M. Ke, S. Peng, and Z. Liu, AIP Advances **6**, 085007 (2016).

[70] J. L. Garcia-Pomar, A. Cortija, and M. Nieto-Vesperinas, Phys. Rev. Lett. **100**, 236801 (2008).